\documentstyle[12pt]{article}
    \newcommand{\be}{\begin{equation}}
    \newcommand{\ee}{\end{equation}}
    \def\n{\noindent}
    \catcode `\@=11
    \catcode `\@=12

    \title{\bf\huge Dual spacetimes, Mach's Principle and topological 
defects}
    
    \author{Naresh Dadhich\thanks{E-mail : nkd@iucaa.ernet.in} \\
    {\sl Inter-University Centre for Astronomy \& Astrophysics,}\\
    {\sl Post Bag 4, Ganeshkhind, Pune - 411 007, India.} 
    } 
    \date{}
    \begin{document}
    \maketitle
    
     \begin{abstract}

    By resolving the  Riemann curvature relative to a unit timelike vector 
   into electric and magnetic parts, we define a duality transformation 
which interchanges active and passive electric parts. It implies interchange 
of roles of Ricci and Einstein curvatures. Further by modifying the 
vacuum/flat equation we construct spacetimes dual to the 
   Schwarzschild solution and flat spacetime. The dual spacetimes describe the 
   original spacetimes with global monopole charge and global texture. The 
   duality so defined is thus intimately related to the topological defects 
and also renders the Schwarzschild field asymptotically non-flat which 
augurs well with Mach's Principle.  
    \end{abstract}

    \n PACS numbers : 04.20,04.60,98.80Hw

    \newpage 
    
\section{Introduction}

   In analogy with  the  electromagnetic  field, it is possible to resolve 
the gravitational field; i.e. Riemann curvature tensor  into electric and 
magnetic parts relative to  a 
    unit  timelike  vector [1-2]. In general, a field is produced by charge 
(source) and its menifestation when charge is stationary is termed as 
electric and magnetic when it is moving. Electromagnetic field is the 
primary example of this general feature, which is true for any 
classical field. In 
gravitation, unlike other fields, charge is also of two kinds. In 
addition to the usual charge in terms of non-gravitational energy, 
gravitational field energy also has charge. Thus electric part would 
also be of two kinds corresponding to the two kinds of charge, which we 
term as active and passive. 
    
  The Einstein vacuum   
    equation, written in terms of electric and magnetic parts is symmetric in 
    active and passive electric parts. We define the duality relation as 
    interchange of active and passive electric parts. Then it turns out that 
    the Ricci and the Einstein tensors are dual of each-other. That is, the 
    non-vacuum equation will in general distinguish between active and passive 
    parts and we could have solutions that are dual of each-other [3]. In 
    particular it follows that perfect fluid spacetimes with the equation of 
    states, $\rho - 3p = 0$ and $\rho + p = 0$ are self dual ($\bigwedge 
\rightarrow -\bigwedge$) while the 
    stiff fluid is dual to dust.

  The question is, can we obtain a dual to a 
    vacuum soltuion? Since the equation is 
    symmetric in active and passive parts, it would remain invariant under 
    the duality transformation. However it turns out that in obtaining 
the well-known black hole solutions not all of the vacuum equations are 
used. In particular, for the Schwarzschild solution the equation $R_{00} = 
0$ in the standard curvature coordinates is implied by the rest of the 
equations. If we tamper this equation, the Schwarzschild solution 
would remain undisturbed for the rest of the set will determine it 
completely. However this modification, which does not 
affect the vacuum solution, breaks the symmetry between active and 
passive electric parts leading to non-invariance of the modified equation 
under the duality transformation. Now we can have solution dual to 
vacuum which is different. This is 
    precisely what happens for the Schwarzschild solution. \\
    
     The Schwarzschild is the unique spherically symmetric vacuum 
    solution, which means it characterizes vacuum for spherical symmetry. It 
    is true that not all the equations are used in getting to the solution. 
    In fact it turns out that ultimately the equations reduce to the Laplace 
    equation and its first integral [4-5]. That means the Laplace equation 
becomes free as it would be implied by its first integral equation. 
Without disturbing the Schwarzschild solution we could introduce some energy 
density on the right which would be wiped out by the other 
    equations. The modified equation 
    would turn out to be not invariant under the duality transformation, yet 
    however it admits the Schwarzschild solution as the unique solution. 
   Now the dual set of equations also admits the unique solution, which 
   could be interpreted as representing the Schwarzschild particle with 
   global monopole charge [6]. The static black hole with and without global 
   monople charge are hence dual of each-other. \\
   
   \n Similarly it turns out that flat spacetime could as well be characterized 
   by a duality non-invariant form of the equation. The static solution of the 
   dual equation will represent massless global monopole (putting the 
   Schwarzschild mass zero in the above solution) and the non-static 
   homogeneous solution will give the FRW metric with the equation of state 
   $\rho + 3p =0$, which is the characterizing property of global texture 
   [7-8]. The former could as well be looked upon as spacetime of uniform 
   relativistic gravitational potential [4-5]. Global monopoles and textures 
   are stable topological defects which are produced in phase transitions in 
   the early universe when global symmetry is spontaneously broken [7-10]. In 
   particular a global monopole is produced when the global
    $O(3)$ symmetry is broken into $U(1)$. A solution for a  Schwarzschild  
   particle 
    with  global  monopole charge has been obtained by  Barriola  and 
    Vilenkin  [6]. It therefore follows  
   that the Schwarzschild and the Barriola-Vilenkin solutions are related 
  through the duality transformation. They are dual  of  each-
    other.  Like  the  Schwarzschild solution,  the  global  monopole 
    solution is  also unique. Applications  to  cosmology and properties  
   of  global  monopoles 
    [10-14] and of global textures [7-8,11,15-19] have been studied by 
    several authors. What dual solution signifies is restoration of gauge 
freedom of choosing zero of relativistic potential which was not 
permitted by the vacuum equation that implied asymptotic flatness. This 
means that the dual solution breaks asymptotic flatness of the 
Schwazschild filed without altering its basic physical character. The  
relativistic potential is now given by $\phi = k -M/r$ instead of $\phi = 
-M/r$. This is precisely what is required to make the Schwarzschild field 
consistent with Mach's principle. The constant $k$ brings in the 
information of the rest of the Universe, say for solar system moving
towards the great attractor [20]. The important difference between the 
Newtonian and relativistic understanding of the problem is that constant 
$k$ produces non-zero curvature and hence has non-trivial physical 
meaning. This is the most harmless way of making the field of an isolated 
body consistent with Mach's principle. \\
    
 \n In sec. 2, we shall give the electromagnetic decomposition of the 
Riemann curvature, followeed by the duality transformation and dual 
spacetimes in sec. 3 and concluded  with discussion in sec. 4.

\section{Elctromagnetic decomposition}

   \n We resolve the  Riemann  curvature tensor relative to a unit timelike 
vector [1-2] as follows :
    
    \be
    E_{ac} = R_{abcd} u^b u^d,  \tilde E_{ac} = *R*_{abcd} u^b u^d
    \ee
    
    \be
    H_{ac} = *R_{abcd} u^b u^d = H_{(ac)} - H_{[ac]}
    \ee
    
    \n where
    
    \be
    H_{(ac)} = *C_{abcd} u^b u^d
    \ee
    
    \be
    H_{[ac]} = \frac{1}{2} \eta_{abce} R^e_d u^b u^d.
    \ee
    
    \n Here $c_{abcd}$  is  the  Weyl conformal  curvature, $\eta_{abcd}$
    is  the  4-dimensional volume element. $E_{ab} = E_{ba}, {\tilde E}_{ab}
    = {\tilde E}_{ba}, (E_{ab}, {\tilde E}_{ab}, H_{ab})
    u^b = 0,~ H= H^a_a = 0$ and $u^a u_a = 1$. 
    The Ricci tensor could then be written as
    
    \be
    R_{ab} = E_{ab} + {\tilde E}-{ab} + (E + {\tilde E}) u_a u_b -
    {\tilde E } g_{ab} + \frac{1}{2} H^{mn} u^c (\eta_{acmn} u_b + 
\eta_{bcmn} u_a)
    \ee
    
    \n where $E = E^a_a$        and  $\tilde E = \tilde E^a_a$. 
    It may be noted that $E = (\tilde E + \frac{1}{2} T)/2$            defines 
    the  gravitational  charge density  while ${\tilde E}= - T_{ab}
    u^a u^b$            defines  the 
    energy density relative to the unit timelike vector $u^a$.   \\

   \section{Duality transformation and dual spacetimes}

   \n The vacuum equation, $R_{ab} = 0$ is in general equivalent to  
    
    \be
    E ~ or ~ {\tilde E} = 0,~ H_{[ab]} = 0 = E_{ab} + {\tilde E}_{ab}
    \ee
    
  \n which is symmetric in $E_{ab}$ and ${\tilde E}_{ab}$.\\

    We define the duality transformation as 
   
     \be
     E_{ab} \longleftrightarrow {\tilde E}_{ab}, ~H_{[ab]} = H_{[ab]}.
     \ee
   
   \n Thus the vacuum eqaution (6) is invariant under the duality 
   transformation (7). From eqn. (1) it is clear that the duality 
   transformation would map the Ricci tensor 
   into the Einstein tensor and vice-versa. This is because contraction of 
   Riemann is Ricci while of its double dual is Einstein.  \\   
   
   \n Consider the spherically symmetric metric,
    
    \be
    ds^2 = c^2(r,t) dt^2 - a^2(r,t) dr^2 - r^2 (d \theta^2 + sin^2 \theta
    d \varphi^2).
    \ee
    
    \n The natural choice for the resolving vector in this case is of 
   course it being hypersurface orthogonal, pointing along the $t$-line. 
 From   eqn. (6), $H_{[ab]} = 0$ and $E^2_2 + {\tilde E}^2_2 = 0$ lead 
   to $ac = 1$ (for this, no boundary condition of asymptotic flatness need 
   be used). Now ${\tilde E} = 0$ gives $a = (1-2M/r)^{-1/2}$, which is the 
   Schwarzschild solution. Note that we did not need to use the remaining 
   equation and  $E^1_1 + {\tilde E}^1_1 = 0$, it is  hence free and is
   implied by the rest. Without affecting the Schwarzschild solution, we 
   can introduce some distribution in the 1-direction. \\      

   \n We hence write the alternate equation as
    
    \be
    H_{[ab]} = 0 = {\tilde E},~ E_{ab} + {\tilde E}_{ab}
    = \lambda w_a w_b
    \ee
    
    \n where $\lambda$ is a scalar and $w_a$ is a spacelike unit vector 
along $4$-acceleration. It is clear that it will also 
   admit the Schwarzschild solution as the general solution, and it determines
   $\lambda = 0$. That is for spherical symmetry the above alternate 
   equation also characterizes vacuum, because the Schwarzschild solution is 
   unique. \\
   
    \n  Let us now employ the duality  transformation (7) to the above  
   equation (9) to write  
    
    \be
    H_{[ab]} = 0 = E,~ E_{ab} + {\tilde E}_{ab} = \lambda w_a w_b.
    \ee

    \n Its general solution for the metric (8) is given by 
    
    \be
    c = a^{-1} = (1 - 2k - \frac{2M}{r})^{1/2}.
    \ee

    \n This is the Barriola-Vilenkin solution [6] for the  Schwarzschild 
    particle with global monopole charge, $\sqrt {2k}$. Again we shall 
      have $ac = 1$ and $E=0$ will then yield $c = (1-2k - 2M/r)^{1/2}$ 
      and $\lambda = 2k/r^2$. This  has  non-zero stresses given by
    
    \be
    T^0_0 = T^1_1 = \frac{2k}{r^2}.
    \ee

    \n A global  monopole    is    described    by    a    triplet    scalar,
     $\psi^a (r) = \eta f(r) x^a/r, x^a x^a = r^2$,
    which  through  the usual Lagrangian  generates  energy-momentum 
    distribution at large distance from the core precisely of the  form 
    given  above  in (12) [6].  Like the Schwarzschild solution the 
   monopole solution (11) is also the unique solution of eqn.(10). \\ 
 
    \n If we translate eqns. (9) and (10) in terms of the familiar Ricci 
 components, they would read as
 
    \be
    R^0_0 = R^1_1 = {\lambda},   R^2_2 = 0 = R_{01}
    \ee
 
    \n and
 
    \be
    R^0_0 = R^1_1 = 0 = R_{01},  R^2_2 = {\lambda}.
     \ee
 
     \n For the metric (8), we shall then have $ac = 1$ and $c^2 = f(r) 
      =1 + 2 \phi$, say, and 

     \be
     R^0_0 = - \bigtriangledown^2 \phi 
     \ee

     \be
     R^2_2 = - \frac {2}{r^2} (r \phi)^{\prime} 
     \ee 

    \n Now the set (13) integrates to give ${\phi} = - M/r$ and 
 ${\lambda} = 0$, which is the Schwarzschild solution while (14) will 
 give ${\phi} = - k - M/r$ and ${\lambda} = 2k/{r}^2$, the 
 Schwarzschild with global monopole charge. Thus global monopole owes 
 its existence to the constant $k$, appearing in the solution of the 
 usual Laplace equation implied by eqns. (14) and (15). It defines a 
 pure gauge for the Newtonian theory, which could be chosen freely, 
 while the Einstein vacuum equation determines it to be zero. For the 
 dual-vacuum equation (14), it is free like the Newtonian case but it 
 produces non-zero curvature and hence would represent non-trivial 
 physical and dynamical effect (see $R^2_2 = - 2k/{r}^2 \neq 0$ 
 unless $k = 0$). This is the crucial difference between the 
 Newtonian theory and GR in relation to this problem, that the latter 
 determines the relativistic potential ${\phi}$ absolutely, vanishing 
 only at infinity. This freedom is restored in the dual-vacuum equation, 
 of course at the cost of introducing stresses that represent a global 
 monopole charge. The uniform potential would hence represent a massless 
 global monopole ($M = 0$ in the solution (11)), which is solely 
 supported by the passive part of electric field. It has been argued and 
 demonstrated [5] that it is the non-linear aspect of the field (which 
 incorporates interaction of gravitational field energy density) that 
 produces space-curvatures and consequently the passive electric part. It 
 is important to note that the relativistic potential ${\phi}$ plays the 
 dual role of the Newtonian potential as well as the non-Newtonian role 
 of producing curvature in space. The latter aspect persisits even when 
 potential is constant different from zero. It is the dual-vacuum 
 equation that uncovers this aspect of the field. \\
   
     \n On the other hand, flat spacetime could also in alternative form be 
   characterized by 
    \be
    {\tilde E}_{ab} = 0 = H_{[ab]}, E_{ab} = \lambda w_a w_b
    \ee
    
    \n leading to $c=a=1$, and implying ${\lambda} = 0$ . Its dual will be 
    
    \be
    E_{ab} = 0 = H_{[ab]}, {\tilde E}_{ab} = \lambda w_a w_b
    \ee

    \n yielding the general solution,
    
    \be
    c^{\prime} = a^{\prime} = 0 \Longrightarrow c=1, a = const. = (1-2k)^{-1/2}
    \ee

    \n which is non-flat and represents a global monopole of zero  mass, 
    as it follows from the solution (11) when $M=0$. This is also the 
 uniform relativistic potential solution. \\
  
    \n Further it is known that the equation of state $\rho + 3p = 0$ 
which means $E=0$, characterizes global 
    texture  [7,19]. That  is, the necessary condition  for spacetimes of 
   topological defects;  global  textures  and monopoles is   $E = 0$. 
Like the uniform potential spacetime, it can also be shown that the 
global texture spacetime is dual to flat spacetime.  In the above 
eqns (13) and (14), replace $w_a w_b$ by the projection tensor $h_{ab} = 
g_{ab} - u_a u_b$. Then   
   non-static homogeneous solution of the dual-flat equation (14) is the FRW 
   metric with $\rho +3p = 0$, which  
   determines  the scale factor $S(t) = \alpha t + \beta, $ and $\rho = 3
    (\alpha^2 +$ k) $/ (\alpha t + \beta)^2, $ k $= \pm 1, 0$. This is also 
   the unique non-ststic homogeneous solution. The general solutions of the 
   dual-flat equation are thus the massless global monopole (uniform 
   potential) spacetime in the static case and the global texture spacetime 
   in the non-static homogeneous case. Thus they are dual to flat 
   spacetime.  \\
     
    \n It  turns  out  that spacetimes  with $E=0$  
    can  be 
    generated   by  considering  a  hypersurface   in   5-dimensional 
    Minkowski space defined, for example, by 
    
    \be
    t^2 - x^2_1 - x^2_2 - x^2_3 - x^2_4 = k^2 (t^2 - x^2_1 - x^2_2 - x^2_3)
    \ee

    \n which consequently leads to the metric
    
    \be
    ds^2 = k^2 dT^2 - T^2 [d \chi^2 + sinh^2 \chi (d \theta^2 + sin^2 \theta 
    d \varphi^2)].
    \ee

    \n Here $T^2 = t^2 - x^2_1 - x^2_2 - x^2_3 $ and $\rho = 3(1-k^2)/k^2 T^2$.                               The above construction  will  
    generate spacetimes of global monopole, cosmic strings (and their 
    homogeneous versions as well), and global texture-like  depending 
    upon  the  dimension  and character of the  hypersurface. Of 
    course, $E=0$   always; i.e.  zero   gravitational   mass 
    [11]. The  trace of active part measures the gravitational  charge 
    density,   responsible  for  focussing  of  congruence   in   the 
    Raychaudhuri  equation  [21]. The topological  defects are thus 
   characterized by vanishing of focussing density (tracelessness of active 
   part). \\  
    
     \n Application of the duality transformation, apart from vacuum/flat 
    case  considered  here, has been considered for  fluid  spacetime 
    [3]. The duality transformation could similarly be considered for    
    eletrovac equation including the $\Lambda$-term. Here the analogue
    of the master equation (10) is
    
    \be
    H_{[ab]} = 0, E = \Lambda - \frac{Q^2}{2r^4}, ~ E^b_a + 
    {\tilde E}^b_a = (- \frac{Q^2}{r^4} + \lambda) w_a w^b
    \ee
    
    \n which has the general solution $c^2 = a^{-2} = (1-2k-2M/r
    + Q^2/2r^2 - \Lambda r^2/3)$ and $\lambda = 2k/r^2$. The
    analogue of eqn. (6) will have ${\tilde E} = - \Lambda - Q^2/2r^4$
    instead of $E$ in (20). Thus the duality transformation works in general
    for a charged particle in the de Sitter universe. Similarly spacetime 
dual to the NUT solution has been obtained [22]. In the case of the Kerr 
solution it turns out, in contrast to 
others, that dual solution is not unique. The dual equation admits two 
distinct solutions which include the original Kerr solution [23]. \\

 \section{Discussion}

 \n First of all let us try to get some physical feel of active, passive 
and magnetic parts. For a  canonical resolution relative to a hypersurface 
orthogonal unit 
timelike vector, it follows that $E_{ab}$ would refer to the 
curvature components $R_{0a0a}$, $\tilde E_{ab}$ to $R_{abab}$ and 
$H_{ab}$ to $R_{0aab}$. With reference to the spherically 
symmetric metric (8), it can be easily seen that the  active 
part is crucially anchored onto the Newtonian potential 
appearing in $g_{00} = 1 + 2\phi$, 
while the passive part to the relativistic potential, $g_{11}  = -
(1 + 2\phi)^{-1}$. Note that in obtaining the 
Schwarzschild solution we ultimately solve the Laplace equation, which 
does not take into account contribution of gravitational filed energy as 
source. It can be shown that contribution of gravitational field energy 
goes into curving the space through $g_{11} \ne 1$ leaving the Laplace 
equation undisturbed [4-5]. Thus passive part is created by the field 
energy while the active by non-gravitational energy distribution. The 
magnetic part would as expected be due to motion of sources. \\

\n Under the duality transformation, the vacuum equation remains 
invariant leading to to the same solutions, but the Weyl tensor changes 
sign which would mean $GM \rightarrow -GM$. This happens because 
active part is produced by positive non-gravitational energy while 
passive part by negative field energy, and the interchange of active and 
passive would therefore imply interchange of positive energy and negative 
field energy [2]. \\

 \n Consider the Maxwell like duality $E\rightarrow H, H \rightarrow -E$ 
as given by
\be
E_{ab}\rightarrow H_{ab}, ~H_{ab}\rightarrow - \tilde E_{ab},~~~~
\tilde E_{ab}\rightarrow - E_{ab}
\ee
 
\n This implies $E = 0, H_{[ab]} = 0, E_{ab} + \tilde E_{ab} = 0$ which 
is the vacuum equation (6) and keeps the Einsten action invariant because 
$R = 2(E - \tilde E)$. This 
is a remarkable result indicating that vacuum equation is implied by the 
duality symmetry of the action [2]. Note also that duality transformation 
of the action does not permit the cosmological constant which could however 
be brought in as matter with the specific equation of state. This 
result is similar to the well-known property of GR that equation of motion 
for free particle is
contained in the field equation.\\ 

\n The duality transformation (7) introduces in most harmless manner a 
global 
monopole in the Schwarzschild black hole which amounts to breaking the 
asymptotic flatness. The latter is a necessary requirement for the field  
to be consistent with Mach's principle at the very elementary level. In 
essence, it is obtained by simply retaining the constant of integration 
in the solution of the Laplace equation. Thus it makes no difference at 
the Newtonian level and hence its contribution is purely  relativistic.\\
   
   \n The most general duality-invariant expression consisting of the Ricci 
   and the metric is $R{^a_b} - ({\frac{R}{4}} - {\Lambda}){g{^a_b}}$. 
 This,  without 
   $\Lambda$ equal to zero would be the equation for gravitational 
 instanton, which 
   follows from the $R^2$-action. The instanton action and the field equation 
  are duality-invariant. They are also conformally invariant as well. As a 
   matter 
   of fact conformal invariance singles out the $R^2$-instanton action. That 
   means the conformal invariance includes the duality invariance, while the 
   duality invariance of the Palatini action with the condition that the 
   resulting equation be valid for all values of $R$ would lead to the 
   conformal invariance [24-25]. The simplest and well-known instanton 
   solution is the de Sitter spacetime. Here the duality only leads to 
 the anti-de Sitter. \\

      \n {\bf Acknowledgement  :}  I have pleasure in thanking Jose  Senovilla, 
 and  LK Patel and Ramesh  Tikekar for useful discussions. Above all
it is a matter of great pleasure and privilege to having known Jayant
and worked with him closely in setting up IUCAA. With deep affection
and feeling I dedicate this work to him on his completing 60.\\


\newpage
    \begin{thebibliography}{99}
    
    \bibitem{} M.A.G. Bonilla and J.M.M. Senovilla (1997) Gen.Rel.Grav. {\bf 29},91.
    
    \bibitem{} N.  Dadhich,  On gravo-electromagnetic duality in general 
relativity, submitted.
   
    \bibitem{} N.   Dadhich,   L.K.   Patel  and   R.   Tikekar   (1998) 
   Class. Quantum Grav. {\bf 15}, L27. 
    
     \bibitem{} N. Dadhich (1995) GR-14 Abstracts, A.98.
    
    \bibitem{} N.. Dadhich (1997) On the Schwarzschild field, gr-qc/9704068.
   
    \bibitem{} M. Barriola and A. Vilenkin (1989) Phys.Rev.Lett. {\bf 63},341.
   
    \bibitem{} R.L. Davis (1987) Phys.Rev. {\bf D35},3705.
   
    \bibitem{} N. Turok (1989) Phys.Rev.Lett. {\bf 63}, 2625.
    
    \bibitem{} T.W.B. Kibble (1979) J.Phys. {\bf A9},1347.
    
    \bibitem{} A. Vilenkin (1985) Phys.Rep. {\bf 121},263.
    
    \bibitem{} N. Dadhich and K. Narayan (1998) Gen. Rel. Grav. 
{\bf 30}, L1133.
    
    \bibitem{} D. Harari and C. Lousto (1990) Phys.Rev. {\bf D42},2626.
    
    \bibitem{} G.W. Gibbons, M.E. Oritz, F. Ruiz Ruiz and T.M. Samols  (1992) 
    Nucl.Phys. {\bf B385}, 127.
    
    \bibitem{} N. Dadhich, K. Narayan and U.A. Yajnik (1998) Pramana {\bf 
50}, 307.
    
    \bibitem{} R.J.  Gott  III  and  M.J.  Rees  (1987)   Mon.Not.R.Astr.Soc. 
    {\bf 227}, 453.
    
    \bibitem{} E.W. Kobl (1989) Ap.J. {\bf 344}, 543.
    
    \bibitem{} M Kamionkowski and N. Toumbas (1996) Phys.Rev.Lett. {\bf 77}, 587.
    
    \bibitem{} V. Sahni (1991) Phys.Rev. {\bf D43}, R301.
    
    \bibitem{} D. Notzold (1991) Phys.Rev. {\bf D43}, R961.

    \bibitem{} D. V. Ahluwalia (1998) Mod.. Phys. Lett. {\bf A13}, 1397.

    \bibitem{} A.K. Raychaudhuri (1955) Phys.Rev. {\bf 90}, 1123.
   
     \bibitem{} M. Nouri-Zonoz, N. Dadhich and D. Lynden-Bell (1999) 
 Class. Quantum Grav. {\bf 16}, 1.

     \bibitem{} N. Dadhich and L. K. Patel (1999) Gravo-electric dual of 
the Kerr solution, submitted.
   
     \bibitem{} K.B. Marathe and G. Martucci (1998) Nuovo Cim. {\bf 
B113}, 1175.
   
     \bibitem{} N. Dadhich and K.B. Marathe (1998) Electromagnetic 
   resolution of curvature and gravitational instantons, submitted to 
Nuovo Cim.
   
      \end{thebibliography}
    \end{document}